\documentclass[submission,copyright,creativecommons]{eptcs}
\providecommand{\event}{GASCom 2024} 

\usepackage{iftex}
\usepackage{amssymb,amsmath}
\usepackage{wasysym}
\usepackage{graphicx}
\usepackage{epsfig}
\usepackage{latexsym}
\usepackage[english]{babel}
\usepackage[utf8]{inputenc}
\usepackage{csquotes}
\usepackage{algorithm} 
\usepackage{algpseudocode}
\usepackage{tikz}
\usetikzlibrary{arrows,shapes.geometric,positioning,calc}
\usepackage[colorinlistoftodos]{todonotes}
\usepackage{changepage}
\usepackage{subfig}
\usepackage{soul}
\usepackage{hyperref}
\usepackage{todonotes}
\usepackage{enumerate}

\newtheorem{thm}{Theorem}
\newtheorem{prop}[thm]{Proposition}
\newtheorem{cor}[thm]{Corollary}
\newtheorem{lem}[thm]{Lemma}

\newtheorem{exa}[thm]{Example}

\newsavebox{\qedB}

\newcommand{\bprop}{\begin{prop}}
\newcommand{\eprop}{\end{prop}}
\newcommand{\bcor}{\begin{cor}}
\newcommand{\ecor}{\end{cor}}
\newcommand{\blem}{\begin{lem}}
\newcommand{\elem}{\end{lem}}

\newcommand{\pset}{[\![b]\!]}
\newcommand{\epset}{\exp{[\![b]\!]}}

\ifpdf
  \usepackage{underscore}         
  \usepackage[T1]{fontenc}        
\else
  \usepackage{breakurl}           
\fi

\title{On the Orthogonality of Generalized Pattern Sequences}
\author{Shuo Li
\institute{Department of Mathematics \& Statistics\\ The University of Winnipeg\\ Winnipeg, Canada}
\email{sh.li@uwinnipeg.ca}}
\def\titlerunning{On the Orthogonality of Generalized Pattern Sequences}
\def\authorrunning{S. Li}

\begin{document}
\maketitle

\begin{abstract}
The partial sums of integer sequences that count the occurrences of a specific pattern in the binary expansion of positive integers have been investigated by different authors since the 1950s. In this note, we introduce generalized pattern sequences, which count the occurrences of a finite number of different patterns in the expansion of positive integers in any integer base, and analyze their partial sums.

\end{abstract}

\section{Introduction, definitions and notation}

Let $b$ be a positive integer larger than $1$. Define $[\![b]\!]=\left\{0,1,2, \cdots ,b-1\right\}$ and $[\![b]\!]^*$ as the set of finite words composed of letters from $[\![b]\!]$. A finite weighted subset $S$ of $[\![b]\!]^*$ is a set of the form
$$\{(n_{S,w}, w)| n_{S,w} \in \mathbf{R}, w \in [\![b]\!]^*\},$$ such that 
$|\{w |\; n_{S,w} \neq 0\}| < \infty.$ For any $w \in [\![b]\!]^*\setminus[\![1]\!]^*$ and any non-negative integer $n$, let $e_{b,w}(n)$ denote the total number of occurrences of the word $w$ in the {\em $b$-expansion} of $n$. In this article, by $b$-expansion of integers, we mean the canonical $b$-expansion of integers with infinitely many leading zeros. For example, $e_{2,0011}(6)=1$, $e_{2,0011}(51)=2$. For any weighted subset $S$ of $[\![b]\!]^*$ and any non-negative integer $n$, define 
$$e_{b,S}(n)=\sum_{(n_{S,w}, w) \in S}n_{S,w}e_{b,w}(n).$$
For any positive integer $m$ larger than $1$, define $a_{b,m,w}(n)=\exp{\frac{2\pi ie_{b,w}(n)}{m}}$ and $a_{b,m,S}(n)=\exp{\frac{2\pi ie_{b,S}(n)}{m}}$ for all non-negative $n$. Both sequences $(a_{b,m,S}(n))_{n \in \mathbf{N}}$ and $(e_{b,m,S}(n))_{n \in \mathbf{N}}$ are well studied in the literature. The sequences $(e_{b,w}(n))_{n \in \mathbf{N}}$ are called {\em block-counting sequences} and $(e_{b,S}(n))_{n \in \mathbf{N}}$ are called {\em digital sequences} from \cite[Chapter 3.3]{allouche_shallit_2003}. The analytical and combinatorial properties of these sequences in the case of $b=n=2$ have been well studied since Thue. The sequences $(a_{2,2,S}(n))_{n \in \mathbf{N}}$ are called {\em pattern sequences} in~\cite{Zheng2018,KONIECZNY2021,Zheng2021}. For some special examples, the $\pm 1$-Thue-Morse sequence can be defined as $(a_{2,2,1}(n))_{n \in \mathbf{N}}$ (see, for example, \cite[P. 15]{allouche_shallit_2003}, the sequence defined there is actually the $\{0,1\}$-Thue-Morse sequence, the sequence $(a_{2,2,1}(n))_{n \in \mathbf{N}}$ can be obtained by changing $0$ to $1$ and $1$ to $-1$ from the previous sequence) and the $\pm 1$-Rudin-Shapiro sequence can also be defined as $(a_{2,2,11}(n))_{n \in \mathbf{N}}$ (see, for example,\cite[Example 3.3.1]{allouche_shallit_2003}). The asymptotic and combinatorial properties of $(a_{2,2,S}(n))_{n \in \mathbf{N}}$ and $(e_{2,2,S}(n))_{n \in \mathbf{N}}$ are studied in~\cite{allouche91,morton90,DavidW1989,ALLOUCHE1992163,Allouche99,ALLOUCHE20033}.

Let $(f_n)_{n \in \mathbf{N}}$ and $(g_n)_{n \in \mathbf{N}}$ be two real sequences. They are {\em orthogonal} if $$\lim_{N \to \infty}\frac{1}{N}\sum_{n=0}^Nf_ng_n=0.$$ The orthogonality between $(a_{2,2,w}(n))_{n \in \mathbf{N}}$ and periodic sequences was first studied by Rudin and Shapiro; they proved separately in \cite{Rudin} and~\cite{Shapiro} that 
$$\max_{0 \leq \theta <1}|\frac{1}{N}\sum_{n=0}^Na_{2,2,11}(n)e^{2 \pi i n \theta}|\leq C\sqrt{N}, N \geq 1.$$ A similar result involving $(a_{2,2,1}(n))_{n \in \mathbf{N}}$ was obtained by Gelfond in~\cite{Gelfond1968}. Moreover, from \cite[Proposition 3.1]{cateland} and~\cite[Theorem 16.1.5]{allouche_shallit_2003}, the sequences of the form $(a_{2,2,S}(n))_{n \in \mathbf{N}}$ are actually in the class of {\em automatic sequences} (see~\cite{allouche_shallit_2003}) and the sequences of the form $(e^{2 \pi i n \theta})_{n \in \mathbf{N}}$ are {\em multiplicative sequences} (see~\cite{Fran19}). The orthogonality between automatic sequences and multiplicative sequences was studied in~\cite{Fran19}. In a series of recent articles \cite{Zheng2018,KONIECZNY2021,Zheng2021}, the correlations of $(a_{2,2,S}(n))_{n \in \mathbf{N}}$ were studied from a viewpoint of dynamical systems. The focus of this paper is two-fold. First, we study the orthogonality among the sequences of the form $(a_{b,m,S}(n))_{n \in \mathbf{N}}$. We later prove that it amounts to study the partial sums of $(a_{b,m,S}(n))_{n \in \mathbf{N}}$. Second, we generalize the result in \cite{DavidW1989} concerning the sequences  $(a_{2,2,w}(n))_{n \in \mathbf{N}}$ to $(a_{b,m,S}(n))_{n \in \mathbf{N}}$ for arbitrary $b$, $m$ and $S$ by using a recent result on the combinatorial structure of $(a_{b,m,w}(n))_{n \in \mathbf{N}}$ introduced in \cite{ABRAM}. We give a necessary and sufficient condition for 

\[\lim_{N \to \infty}\frac{1}{N}\sum_{n=0}^N a_{b,m,S}(n)=0. \tag{$\star$}\]
The main results are announced in Theorem \ref{thm:subst} and Theorem \ref{thm:sum}.

\section{Window functions and $(a_{b,m,S}(n))_{n \in \mathbf{N}}$}
\label{sec:wind}


A finite weighted subset $S$ of $\pset^*$ is called {\em proper} if $n_{S,w} \neq 0$ implies $w$ does not have leading zeros. For any two finite weighted subsets $S_1, S_2$ of $\pset^*$, define 
\[S_1 \oplus S_2=\{(n_{S_1,w}+n_{S_2,w},w)|\; w \in \pset^*\}.\] 
Any finite word $w$ in $[\![b]\!]^*$ is written  $w=w[1]w[2]\cdots w[|w|]$ where  $|w|$ is its  {\em length}.  For later use we denote $w'=w[2]w[3] \cdots w[|w|]$,  and  $(w)_b=\sum_{i=1}^{|w|}w[i]b^{|w|-i}$. \\

Let $\epset=\{e^{\frac{2\pi i n}{b}}|\; n \in \pset\}$ and  $\epset^*=\{e^{\frac{2\pi i n}{b}}|\; n \in \pset\}^*$ be the set of finite words composed of letters from $\epset$. \\

Let us generalize the definition of the {\em window function} in~\cite{ABRAM}. For any integers $b,m$ larger than $1$ and any $w \in \pset^{*}$, let $\alpha^1_w=\frac{(w')_b}{b^{|w|-1}}$ and $\alpha^2_w=\frac{(w')_b+1}{b^{|w|-1}}$. The {\em window function} $\phi_{b,m,w}: \epset^*\to \epset^*$ is  such that for any $v \in  \epset^*$:
$$
\phi_{b,m,w}(v)[j]=
\begin{cases}
e^{\frac{2\pi i}{m}}v[j], & \text{if $\alpha^1_w|v| < j  \leq \alpha^2_w|v|$;}\\
v[j], & \text{otherwise.}
\end{cases}
$$
It is extended to finite weighted subset $S$ of $\pset^*$, by setting 
$$\phi_{b,m,S}=\prod_{(n_{S,w},w) \in S}(\phi_{b,m,w})^{n_{S,n}}.$$

\begin{prop}
\label{prop:multi}
 Let $b,m$ be two integers larger than $1$ and let $S_1, S_2$ be two finite weighted subsets of $\pset^*$. For any non-negative integer $n$, one has
 $$a_{b,m,S_1}(n)a_{b,m,S_2}(n)=a_{b,m, S_1\oplus S_2}(n).$$ 
\end{prop} 

As a corollary, the orthogonality of two generalized pattern sequences is equivalent to the $(\star)$ property of some generalized pattern sequence.

\begin{prop}
\label{prop:change}
 Let $b, m$ be two integers larger than $1$ and let $S$ be a finite weighted subset of $\pset^*$. There exists a finite proper weighted subset $S'$ of $\pset^*$ such that for any non-negative integer $n$, one has
 $a_{b,m,S}(n)=a_{b,m, S'}(n).$
\end{prop} 

An analog of Proposition 3 in~\cite{ABRAM} is stated in our first claimed result:

\begin{thm}
\label{thm:subst}
 Let $b,m$ be two integers larger than $1$ and let $S$ be a finite weighted subset of $\pset^*$. Let $l=\max\{|w|\; | \;n_{S,w} \neq 0\}$. There exist $k-1$ proper weighted subset $S_1,S_2, \ldots, S_{b-1}$ of $\pset^*$ and a sequence $(u_t)_{t\in \mathbf{N}}$ in $\epset^*$ such that:
 \begin{enumerate}
 \item $|u_0|=b^l$;
 \item $u_{t+1}=u_t\phi_{b,m,S_1}(u_t)\phi_{b,m,S_2}(u_t)\cdots\phi_{b,m,S_{p-1}}(u_t)$ for all $t \geq 0$;
\item $(a_{b,m,S}(n))_{n \in \mathbf{N}}= \lim_{t \to \infty}u_t$.
 \end{enumerate}
\end{thm} 

\section{Application}

From Theorem~\ref{thm:subst}, one can associate a generalized pattern sequence $(a_{b,m,S}(n))_{n \in \mathbf{N}}$ to a matrix $M_{b,m,S}$ in the following way:
\begin{enumerate}
 \item let $S_1,S_2, \ldots, S_{b-1}$ and $l$ be the same as in Theorem~\ref{thm:subst};
\item let $V_0$ be the constant sequence of $1$ with a length of $p^l$,  and let $V_k=\phi_{b,m,S_k}(V_0)$ for all $k \in \{1,2,\cdots,b-1\}$;
\item let $M_{b,m,S}\in \mathcal{M}_{p^l \times p^l}(\mathbf{C})$ such that for any integers $1 \leq r \leq b^{l-1}$, $0 \leq s \leq b-1$ and $1 \leq t \leq b$,
$$M_{b,m,S}(x,y)=\begin{cases}
V_s[(r-1)b+t], & \text{if $(x,y)=(sb^{l-1}+r,(r-1)b+t)$;}\\
0, & \text{otherwise}.
\end{cases}$$
\end{enumerate}

\begin{thm}
\label{thm:matr}
 Let $b,m$ be two integers larger than $1$, let $S$ be a finite weighted subset of $\pset^*$ and let $(a_{b,m,S}(n))_{n \in \mathbf{N}}$ and $M_{b,m,S}$ be respectively the associated generalized pattern sequence and the matrix. Let $(A_{b,m,S}(m))_{m \in \mathbf{N}}$ be a sequence of column vectors of dimension $p^l$ such that for any integers $t \geq 0$ and $1 \leq j \leq p^l$, 
 $$A_{b,m,S}(t)(j)=\sum_{n=(j-1)p^{t}}^{jp^{t}-1} a_{b,m,S}(n).$$
 Then for any integer $t \geq 0$, one has
 $$A_{b,m,S}(t+1)=M_{b,m,S}A_{b,m,S}(t).$$
\end{thm}

And we have our second claimed result.

\begin{thm}\label{thm:sum}
 Let $b,m$ be two integers larger than $1$, let $S$ be a finite weighted subset of $\pset^*$. The sequence $(a_{b,m,S}(n))_{n \in \mathbf{N}}$ satisfies the property $(\star)$ if and only if at least one of the following conditions holds:
 \begin{enumerate}
 \item $b$ is not an eigenvalue of $M_{b,m,S}$;
 \item $M_{b,m,S}^{b-1}A_{b,m,S}(0)=\mathbf{0}$, where $A_{b,m,S}(0)$ is as the same as in Theorem~\ref{thm:matr}. 
 \end{enumerate}
\end{thm}
\medskip

\begin{exa}{\rm
Let $b=m=3$ and let $S=\{(n_{S,w},w)|n_{S,w} \in \mathbf{R}, w \in \{0,1\}^*\}$ satisfying
$$n_{S,w}=\begin{cases}
1, &\text{if $w=1,10,12$;}\\
2,& \text{if $w=11,22$;}\\
0, & \text{otherwise.}
\end{cases}$$
From Proposition 3 in~\cite{ABRAM}, the window functions associated to $w=1,10,12,11,22$ are respectively:
$${ \phi_{3,3,1}(v)[j]=
e^{\frac{2\pi i}{q}}v[j],\;\;\;\;\;\;\;\;\;\;\;\;\;\;\;\;\;\;\;\;\;\;\;\;\;\;\;\; \phi_{3,3,12}(v)[j]=
\begin{cases}
e^{\frac{2\pi i}{q}}v[j], & \text{if $\frac{2}{3}|v| < j  \leq |v|$;}\\
v[j], &\text{otherwise.}
\end{cases}, }$$

$${ \phi_{3,3,10}(v)[j]=
\begin{cases}
e^{\frac{2\pi i}{q}}v[j], & \text{if $0< j  \leq \frac{1}{3}|v|$;}\\
v[j], & \text{otherwise.}
\end{cases} \quad
\phi_{3,3,11}(v)[j]=
\begin{cases}
e^{\frac{4\pi i}{q}}v[j], & \text{if $\frac{1}{3}|v| < j  \leq \frac{2}{3}|v|$;}\\
v[j], & \text{otherwise.}
\end{cases},} $$

$${ \phi_{3,3,22}(v)[j]=
\begin{cases}
e^{\frac{4\pi i}{q}}v[j], & \text{if $\frac{2}{3}|v|< j  \leq |v|$;}\\
v[j], & \text{otherwise.}
\end{cases}.} $$
From Theorem~\ref{thm:subst}, one can find $S_1, S_2$ satisfying
$$n_{S_1,w}=\begin{cases}
1, & \text{if $w=1,10,12$;}\\
2, & \text{if $w=11$;}\\
0,& \text{otherwise.}
\end{cases} \quad
n_{S_2,w}=\begin{cases}
2, & \text{if $w=22$;}\\
0, & \text{otherwise.}
\end{cases}$$
Thus,
$${ \phi_{3,3,S_1}(v)[j]=
\begin{cases}
e^{\frac{2\pi i}{q}}v[j], & \text{if $\frac{1}{3}|v| < j  \leq \frac{2}{3}|v|$;}\\
e^{\frac{4\pi i}{q}}v[j], &\text{otherwise.}
\end{cases} \quad
\phi_{3,3,S_2}(v)[j]=
\begin{cases}
e^{\frac{4\pi i}{q}}v[j], &\text{if $\frac{2}{3}|v| < j  \leq |v|$;}\\
v[j], &\text{otherwise.}
\end{cases}. }$$

From Theorem~\ref{thm:subst}, set $u_0=1,1,1,e^{\frac{4\pi i}{3}},1,e^{\frac{4\pi i}{3}},1,1,e^{\frac{4\pi i}{3}}$, and define a sequence of words $(u_n)_{n \in \mathbf{N}}$ such that $u_{m+1}=u_m\phi_{3,3,S_1}(u_m)\phi_{3,3,S_2}(u_m)$ for all integers $m \geq 0$, then $(a_{3,3,S}(n))_{n \in \mathbf{N}}= \lim_{m \to \infty}u_m$.
Moreover, the associated matrix $$M_{3,3,S}=\begin{pmatrix}
1 & 1 & 1 &0&0&0&0&0&0\\
0 & 0 & 0 &1&1&1&0&0&0\\
0 & 0 & 0 &0&0&0&1&1&1\\
e^{\frac{4\pi i}{q}} & e^{\frac{4\pi i}{q}}& e^{\frac{4\pi i}{q}} &0&0&0&0&0&0\\
0 & 0 & 0 &e^{\frac{2\pi i}{q}}&e^{\frac{2\pi i}{q}}&e^{\frac{2\pi i}{q}}&0&0&0\\
0 & 0 & 0 &0&0&0&e^{\frac{4\pi i}{q}}&e^{\frac{4\pi i}{q}}&e^{\frac{4\pi i}{q}}\\
e^{\frac{2\pi i}{q}} & e^{\frac{2\pi i}{q}} & e^{\frac{2\pi i}{q}} &0&0&0&0&0&0\\
0 & 0 & 0 &e^{\frac{2\pi i}{q}}&e^{\frac{2\pi i}{q}}&e^{\frac{2\pi i}{q}}&0&0&0\\
0 & 0 & 0 &0&0&0&e^{\frac{4\pi i}{q}}&e^{\frac{4\pi i}{q}}&e^{\frac{4\pi i}{q}}\\
\end{pmatrix}$$
Since $3$ is not an eigenvalue of $M_{3,3,S}$, we have $$\lim_{N \to \infty}\frac{1}{N}\sum_{n=0}^N a_{3,3,S}(n)=0.$$

}
\end{exa}

\begin{exa}{\rm
Let $b=m=2$ and let $U$ be a finite weighted subset of $\{0,1\}^*$ satisfying
$$n_{U,w}=\begin{cases}
1, & \text{if $w=1,10,11$;}\\
0, & \text{otherwise.}
\end{cases}$$
The associated matrix $$M_{2,2,U}=\begin{pmatrix}
1 & 1 &0&0\\
0 & 0 &1&1\\
1 & 1 & 0 &0\\
0 & 0 &1&1\\
\end{pmatrix}.$$
One can easily verify that $2$ is an eigenvalue of $M_{2,2,U}$. However, since $A_{2,2,U}(0)=(1,-1,1,-1)^t$, and $M_{2,2,U}A_{2,2,U}(0)=\mathbf{0}$, one has
$$\lim_{N \to \infty}\frac{1}{N}\sum_{n=0}^N a_{2,2,U}(n)=0.$$
In fact, one can verify $(a_{2,2,U}(n))_{n \in \mathbf{N}}$ is the $(1,-1)$-periodic sequence.

}
\end{exa}

\begin{exa}{\rm
Let us consider $(a_{3,3,002}(n))_{n \in \mathbf{N}}$. One has, for any non-negative integer $n$, $$\begin{aligned}e_{3,3,002}(n)&=e_{3,3,02}(n)-e_{3,3,102}(n)-e_{3,3,202}(n)\\
&=e_{3,3,2}(n)-e_{3,3,12}(n)-e_{3,3,22}(n)-e_{3,3,102}(n)-e_{3,3,202}(n).\end{aligned}$$
Thus, 
$$a_{3,3,002}(n)=a_{3,3,2}(n)(a_{3,3,12}(n)a_{3,3,22}(n)a_{3,3,102}(n)a_{3,3,202}(n))^{-1}.$$
Define two finite weighted subsets $P,Q$ of $\{0,1,2\}^*$ such that 
$$n_{P,w}=\begin{cases}
-1, & \text{if $w=12,102$;}\\
0, & \text{otherwise.}
\end{cases} \quad
n_{Q,w}=\begin{cases}
1, & \text{if $w=2$;}\\
-1, & \text{if $w=22,202$;}\\
0, & \text{otherwise.}
\end{cases}$$
 From Theorem~\ref{thm:subst}, set $u_0=a_{3,3,002}(0),a_{3,3,002}(1),\cdots,a_{3,3,002}(26)$, and define a sequence of words $(u_n)_{n \in \mathbf{N}}$ such that $u_{m+1}=u_m\phi_{3,3,P}(u_m)\phi_{3,3,Q}(u_m)$ for all integer $m \geq 0$, then $(a_{3,3,002}(n))_{n \in \mathbf{N}}= \lim_{m \to \infty}u_m$.
}
\end{exa}

\bibliographystyle{eptcs}
\bibliography{generic}
\end{document}